\documentclass[aps,prl,twocolumn,groupedaddress]{revtex4-1}
\usepackage{graphicx} 
\begin{document}

%\preprint{}

\title{Tomography of a High-Purity Narrowband Photon From a Transient Atomic Collective Excitation}


\author{A.~MacRae$^{1}$, T.~Brannan$^{1}$, R.~Achal$^{1}$ and A.~I.~Lvovsky$^{1,2}$}
%\thanks{}
\affiliation{$^1$Institute for Quantum Information Science, University of Calgary, Canada}
\affiliation{$^2$Russian Quantum Center, Moscow 121614, Russia}

\date{\today}

\begin{abstract}
We demonstrate efficient heralded generation of high purity narrow-bandwidth single photons from a transient collective spin excitation in a hot atomic vapour cell. Employing optical homodyne tomography, we fully reconstruct the density matrix of the generated photon and observe a Wigner function reaching the zero value without correcting for any inefficiencies. The narrow bandwidth of the photon produced is accompanied by a high generation rate yielding a high spectral brightness. The source is therefore compatible with atomic-based quantum memories as well as other applications in light-atom interfacing. This work paves the way to preparing and measuring arbitrary superposition states of collective atomic excitations.\end{abstract}

\pacs{}
%\keywords{}

\maketitle

%\section{}
%\subsection{}
%\subsubsection{}

The last decade has brought about significant advances in quantum technology of optical states. Combining the techniques of spontaneous parametric down-conversion (SPDC), linear optical processing, postselection on measurement results and optical homodyne tomography has led to a plethora of experiments in which interesting optical states have been prepared and measured \cite{LvovRaymer09}. Similar accomplishments have been demonstrated in the microwave \cite{Hofheinz_09} and trapped ion \cite{Leibfried_05} regimes.

A natural next frontier in quantum state technology is to extend methods of quantum state engineering to collective spin excitations (CSEs) of atomic ensembles. Such extension can find applications in  quantum memory \cite{Lvovsky_09}, long distance quantum communication \cite{DLCZ_01}, quantum logic gates \cite{You_00}, and quantum metrology \cite{Kasevich_11}. In addition, quantum engineering within the Hilbert space of atomic CSEs is of fundamental interest, as it allows one to explore the isomorphism with the Hilbert space of a single electromagnetic mode \cite{Kitagawa_93}. So far, engineering of CSEs has been limited to squeezed spin states \cite{Appel_09, Leroux_10,Fernholz_08} and the single-quantum state \cite{Chou_04,Thompson_06}.

The single CSE quantum can be prepared by heralding on detection of a photon that has undergone Raman scattering from an atomic ensemble, according to the idea of Duan, Lukin, Cirac and Zoller (DLCZ) \cite{DLCZ_01}. The Hamiltonian governing the Raman scattering event is identical to that of SPDC, leading to the production of a two-mode squeezed state of the scattered light and the CSE. While DLCZ utilizes only the first-order term of the evolution under this Hamiltonian, higher-order terms can be used in combination with complex measurements on the scattered optical mode to produce arbitrary quantum CSE states akin to Bimbard \textit{et al.} \cite{Bimbard_10}.

Once the desired collective state has been produced, it needs to be measured. To that end, the readout stage of the DLCZ protocol may  be used, in which the CSE is converted into the optical domian in a manner similar to readout from a quantum optical memory based on electromagnetically-induced transparency \cite{DSP_00}. Full information about the retrieved optical state, and hence about the CSE, can then be acquired using optical homodyne tomography. An alternative technique of performing tomography on atomic CSEs involves off-resonant Faraday interactions \cite{Fernholz_08}.

This outlines an approach to synthesis and measurement of arbitrary quantum states of atomic CSEs. Here we present a proof-of-principle experiment to demonstrate the validity of this approach. We produce a heralded single photon from a transient CSE in an atomic vapor cell. For the first time for a photon from an atomic source, we perform homodyne tomography thereupon, obtaining unprecedented \emph{uncorrected} measurement efficiency of about 50\%, leading to a Wigner function which reaches a zero value at the origin of the phase space. In this way, our experiment completes the toolbox required for complete atomic state engineering.

Aside from this fundamental aspect, our setup can be viewed as a highly-efficient, spectrally bright source of single photons for experiments on interfacing quantum information between light and atoms. In this context, a key requirement is that the photon spectrum be compatible with the atomic transition in both its bandwidth and central frequency. A further crucial figure of merit is the \emph{spectral brightness} measured in the number of photons per second per unit bandwidth. Thus far, the majority of quantum light sources do not perform well in view of these requirements. For example, SPDC-based heralded sources, while featuring a high photon production rate, also exhibit spectral widths on the order of $10^6$ MHz. Cavity \cite{Polzik_07} or waveguide \cite{Halder_08} enhanced SPDC sources have greatly improved the brightness but both are based on filtering a portion of a wider bandwidth and comes at the expense of increased experimental complexity. Single emitters such as atoms in a cavity \cite{Specht_09}, quantum dots \cite{Hostein_09},  and nitrogen-vacancy centres in a diamond \cite{Batalov_08} have provided on-demand photons albeit with low brightness and low collection efficiency. Finally, existing DLCZ-based atomic sources of heralded photons, implemented in free space \cite{Chou_04} and in an optical cavity \cite{Thompson_06}, exhibit sufficiently high spectral brightness but relatively poor efficiencies.

Our source of heralded single photons can be seen as a hybrid of SPDC and DLCZ. It is based on  four-wave mixing (4WM) in hot atomic vapour, which has recently been shown to generate high-quality twin-beam optical squeezing in a large array of modes  \cite{McCormick_07,Boyer_08}. We use this twin-beam state to prepare heralded single photons. Specifically, we utilize a single, strong, continuous laser field that simultaneously pumps both Raman transitions associated with the read and write stages of the DLCZ protocol [Fig.~1(a)]. In this sense, our system is analogous to SPDC: a pair of correlated photons is created simultaneously in the phase-matched directions, which define the \emph{signal} and \emph{idler} channels. However, the physics of our setting is quite different: it utilizes the $\chi^{(3)}$ nonlinearity in atoms near resonance, while SPDC is based on off-resonance $\chi^{(2)}$ in a solid. While SPDC constitutes conversion of a single pump photon into the signal and idler photon, in our process the conversion requires two photons from the pump field. The required third-order optical nonlinearity is sufficiently strong only in the neighbourhood of an atomic resonance. Accordingly, our heralded photon source is intrinsically bright, narrowband and produces photons in a well-defined spatiotemporal mode. While the photons are created off resonance to $^{85}$Rb, they are produced directly on resonance with $^{87}$Rb, and may thus be employed in on- and off-resonant protocols. The  spectral brightness is among the highest observed to date \cite{Thompson_06,Polzik_07,Halder_08}, and its measured efficiency exceeds that of its competitors by about one order of magnitude.

\begin{figure}[h!]
  \centering
\includegraphics[width=90mm]{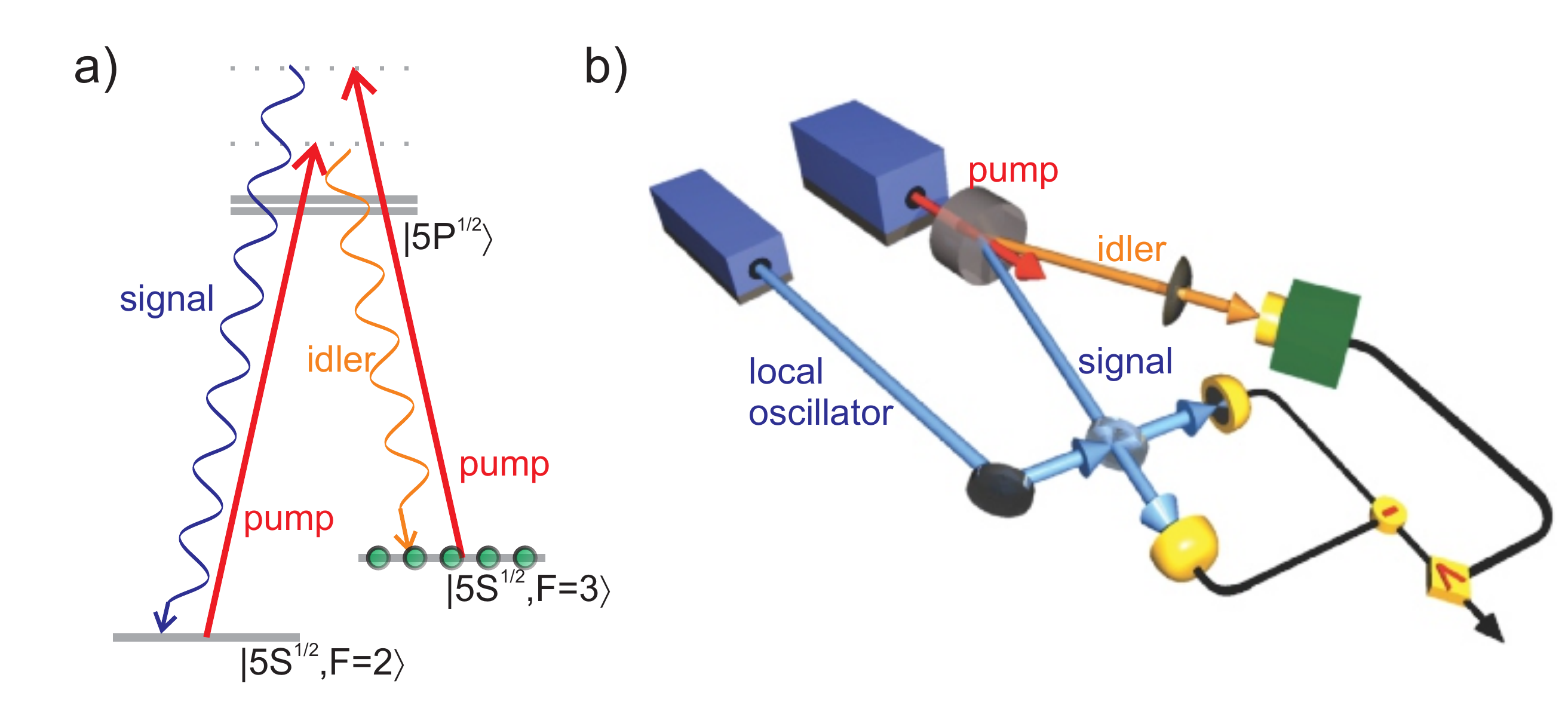}
  \caption{The experiment. (a) The atomic level scheme. Optical pumping prepares most atoms in the $F=3$ hyperfine ground state. Raman scattering of the pump leads to the creation of a collective spin excitation into the $F=2$ state, followed by immediate recycling back into $F=3$. This process results in emission of a correlated pair of the signal and idler photons. (b) The setup. A titanium-sapphire laser is passed through a 12-mm cell containing $^{85}$Rb vapour. The idler photon passes through spatial and spectral filters and is detected on a single photon counting module. Upon a detection event, balanced homodyne detection of the signal photon is performed. To this end, a phase-stable local oscillator is produced by phase-locking a diode laser to the pump laser. }
\end{figure}

The experimental setup is diagrammed in Fig.~1b. The biphotons are generated by passing a single 800-mW pump beam of $e^{-2}$ radius 550 $\mu$m through a 12-mm cell containing $^{85}$Rb vapour at about $100^\circ$C. The laser is tuned near resonance with the D1 transition at $\lambda=795$ nm. The signal and idler photons are created at a frequency difference equal to the hyperfine ground state splitting (3.035 GHz) on each side of the pump frequency and are orthogonally polarized to the pump.

A significant challenge is associated with preventing the strong pump field from contaminating the biphoton measurements. It is addressed in three ways. First, polarization filtering attenuates the pump by 5 orders of magnitude. Second, spatial filtering is implemented: working in a large Fresnel number configuration allows us to achieve phase-matching over a range of angles between the signal and idler photons. We choose to work at an angle of $\pm4.7$ mrad with respect to the pump which is a compromise between phase-matching and spatial separation from the pump. Third, frequency filtering of the idler channel is implemented. To this end, we employ a monolithic spherical Fabry-Perot cavity of a 55-MHz linewidth constructed from a standard lens with high reflectivity coating on each side, tuned by varying its temperature \cite{Cavity_Paper}. The cavity is operated slightly off the 4WM gain peak in order to filter out the spectral region where the idler undergoes Raman absorption, resulting in uncorrelated photon emission (see, for example, \cite{Boyd}). Because of intrinsic spatiotemporal mode filtering associated with homodyne detection, filtering in the idler channel is unnecessary. In experiments employing photon counting, filtering can be accomplished by means of a monolithic filter cavity, as in the idler channel.

Idler photon detection events occur at a rate of $\approx300$ kHz. Upon each such event, the signal channel is measured using a balanced homodyne detector (HD) with a 100-MHz bandwidth \cite{Kumar_11}. The local oscillator for the HD must be phase stable with respect to the pump and must be at the frequency of the generated signal photons. To accomplish this, we lock a diode laser to a frequency $3.035$ GHz blue of the pump beam via an optical phase-lock loop \cite{PLL_09}. The spatial mode of the local oscillator is matched to that of the signal photon by injecting an auxiliary laser beam into the idler channel, which mixed with the pump inside the cell to generate a classical field in the signal mode \cite{Aichele_02}.

A quadrature measurement of the heralded state is obtained from the homodyne current by integrating over its temporal mode function $\psi(t)$:
\begin{equation}\label{tempwf}
Q_\theta = \int_{-\infty}^{\infty} q_\theta(t)\psi(t)dt,
\end{equation}
\noindent where $q_\theta(t)$ is proportional to the instantaneous homodyne detector output photocurrent, and $\theta$ is the phase of the local oscillator with respect to the quantum state, and the time is measured with respect to the trigger event.

Knowledge of the temporal mode (1) of the heralded photon is essential for its efficient reconstruction \cite{Kumar_11}. This mode is determined by the spectral wavefunction of the biphoton produced in 4WM as well as the transmission spectrum of the filter in the idler channel. The former is not known \emph{a priori}. In order to measure the magnitude of the temporal profile, we plot the variance of the HD output current as a function of the delay with respect to the trigger event [Fig.~2(a)]. For points well separated from trigger events, we observe a background thermal state, as expected when a two-mode squeezed state is traced over one of its modes. In a window of approximately $10$ ns from the trigger event we note a marked increase in variance reaching a peak of $2.3$ times the vacuum due to the contribution of the heralded single photon to the quadrature noise. To acquire phase information of the temporal mode, we observe the autocorrelation $\langle q(t_1) q(t_2)\rangle$ of the signal, corresponding to the density matrix of the heralded state in the time domain. The autocorrelation function is of round shape, showing high purity of the temporal mode of the heralded photon. This observation is further confirmed by the Schmidt decomposition of the autocorrelation function. The eigenvector corresponding to the primary eigenvalue yields the temporal mode function as shown in Fig.~2(b). We believe the side lobes of the mode function to be caused by the off-resonant position of the filter cavity.

\begin{figure}[h!]
\centering
\includegraphics[width=85mm]{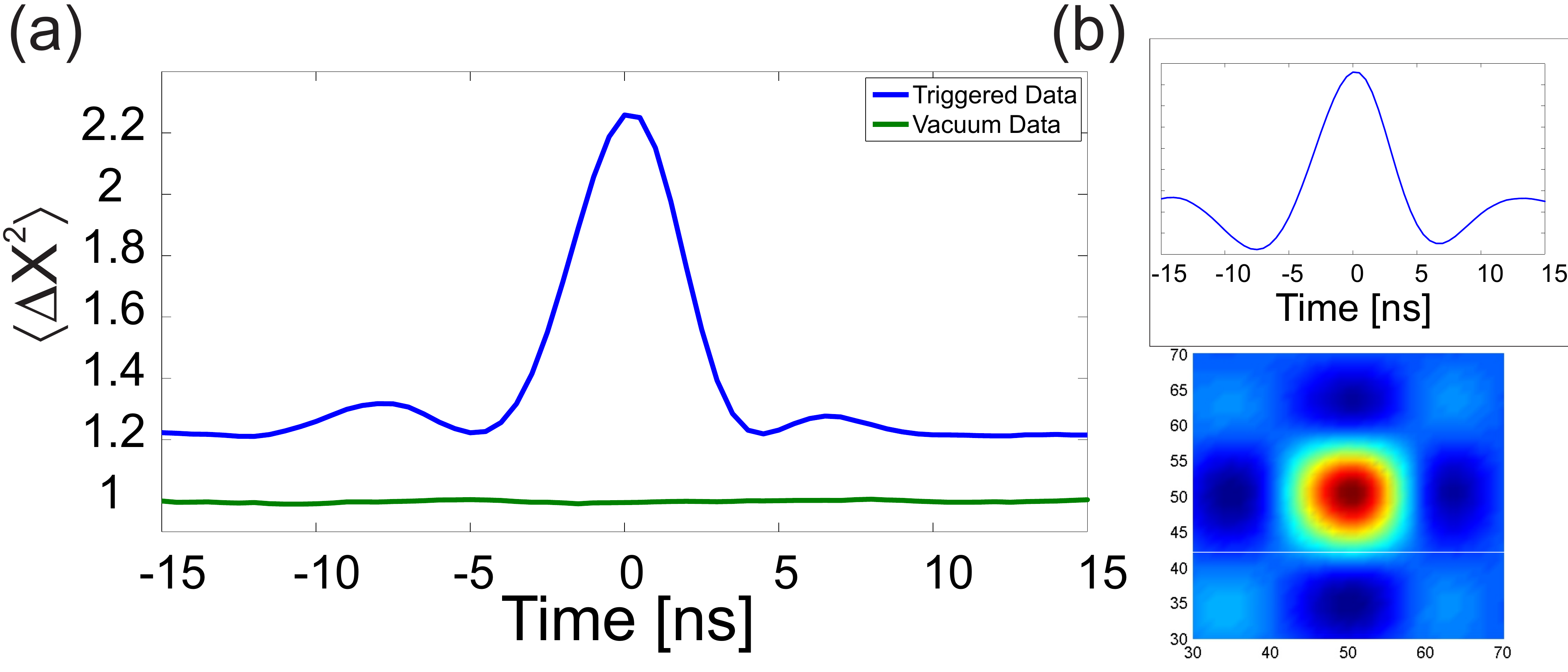}
\caption{Temporal profile of the heralded photon. (a) The magnitude $|\psi(t)|^2$ may be inferred by observing the variance of the homodyne detector photocurrent as a function of delay with respect to the trigger event. (b) In order to infer the phase information of the temporal profile, we construct a matrix corresponding to the autocorrelation of the photocurrent as a function of trigger delay. The eigenvector corresponding to the primary eigenvalue of this matrix (inset) gives the temporal weight function $\psi(t)$.}
\end{figure}

We acquire $10^5$ traces of the homodyne current within a $180$ ns window about the trigger event and integrate each trace over the temporal profile determined above to produce the ensemble of quadrature values. The quadrature statistics exhibit no phase dependence, as expected for a statistical mixture of Fock states. As seen in Fig.~3(a), the observed quadrature probability distribution is highly non-Gaussian and shows a well-defined dip at the origin characteristic of the single-photon state. If the observed state were a perfect single photon, the probability to record a quadrature value of zero would vanish, but this is not the case, mainly due to the admixture of vacuum from losses and mode mismatch.

In order to determine the density matrix of the detected state, we apply an iterative maximum-likelihood algorithm \cite{MaxLik_04,MaxLik_07} obtaining an uncorrected single photon probability of $\rho_{11}=0.488\pm0.002$. The two- and three-photon contributions equal $\rho_{22}=0.069$ an $\rho_{33}=0.019$, respectively. The reconstructed Wigner function is shown in Fig.~3(b) displaying a negative value of $-0.0045\pm0.0025$ at the origin of the phase space because the fraction of the odd-number states (primarily the single-photon state) is higher than that of the even-number states.

\begin{figure}[h!]
  \centering
\includegraphics[width=85mm]{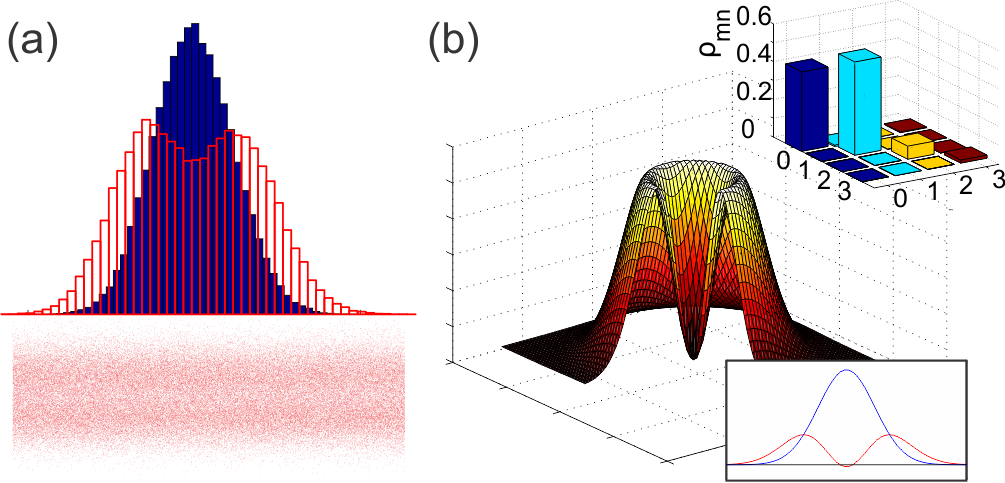}
\caption{Experimental results. (a) The marginal quadrature distribution for the heralded state (red) and the vacuum (solid blue), showing a dip near the origin. The raw set of quadrature samples is shown below. (b) The Wigner function obtained from the reconstructed density matrix (upper inset).  The lower inset shows the cross-section of the Wigner function reaching negative values at the origin.}
\end{figure}

Since the density matrix gives full information of the quantum state, we can calculate common figures of merit for single photon sources. We find that in the uncorrected reconstructed temporal mode the second order correlation of the field is $g^{(2)}(0)\equiv\langle\hat{a}^\dagger\hat{a}^\dagger\hat{a}\hat{a}\rangle/\left(\hat{n}\right)^2=0.51<1$ and the Mandel Q-parameter is $Q=-0.32<0$, evidencing the nonclassical character of the generated state. We note that the difference between the ideal value of $g^{(2)}=0$ and our reported figure is primarily due to the higher-number terms. With lower gain corresponding to a temperature of $69^\circ$C, our two-photon component is negligible and is bounded above only by the statistical uncertainty of our measurement ($0.3\%$) yielding $g^{(2)}(0)\leq0.13$, albeit at the expense of reducing the single-photon efficiency to $21\%$. From the quadrature variance of the triggered signal $\langle\Delta\hat{X}^2_{\text{trig}}\rangle$ and the thermal background $\langle\Delta\hat{X}^2_{\text{bck}}\rangle$, we can also calculate the conditional cross-correlation between the signal and idler channels $g^{(2)}_{\text{si}}=(\langle\Delta\hat{X}^2_{\text{trig}}\rangle-1/2)/(\langle\Delta\hat{X}^2_{\text{bck}}\rangle-1/2)$ \cite{Polzik_07}. At lower temperatures we find $g^{(2)}_{\text{si}}=24.2>1$ at zero delay whereas for higher temperatures we measure $g^{(2)}_{\text{si}}=6.0$.

From the Fourier transform of the temporal mode, we observe the spectral bandwidth to be $39$ MHz, corresponding to a spectral brightness of $7,700$ photons per MHz per second, comparable to the highest values achieved to date \cite{Halder_08}. The fundamental limit to spectral brightness is set by the requirement that the mean photon number of the thermal background state be much less than one, to avoid significant overlap of photons. The maximum count rate must then be much less than the photon bandwidth. In the present work, taking into account the efficiency of the single photon detector, we are operating on the border of this limit.

All imperfections characteristic of heralded single-photon experiments, such as the linear losses, nonzero detection efficiency, imperfect mode matching between the local oscillator and the signal \cite{Lvovsky_01} exist in the present setting as well. Additional imperfections are specific to atomic vapour cells, such as uncorrelated Raman scattering into the signal and idler channels owing to the population exchange of the hyperfine ground states. This decoherence mechanism is partially overcome by strong optical pumping by the driving field but this has the detrimental effect of broadening the emission line width and introducing significant higher-order terms into the two-mode squeezed state of the signal and idler. We therefore expect that further improvements to the single-photon fidelity can be achieved by incorporating a conservative quantity of buffer gas and/or polymer wall coating into the vapour cell. This modification will also lead to further reduction of the photon bandwidth, since the bandwidth of the Raman process decreases with increasing pump Rabi frequency.

To summarize, we have developed a new high spectral brightness, high purity source of heralded single photons which are compatible with atomic quantum memories. The observed uncorrected single-photon efficiency of approximately $50\%$ is an order of magnitude higher than previously reported figures for atomic photons sources. Employing optical homodyne tomography, we reconstructed the density matrix of the heralded state and found the corresponding Wigner function to be negative. Because the photon is obtained from a transient coherent spin excitation of the ground state, the experiment can be viewed as characterization of the quantum state of this excitation, paving the way to a new range of experiments on state engineering of atomic ensembles.

\begin{acknowledgments}

The authors thank M. F\"{o}rtsch for his help in the experiment; A. Marino, J.-M. Wen, and C. Simon for fruitful discussions. We acknowledge financial support from iCore, AITF, NSERC, and CIFAR.
\end{acknowledgments}

%\bibliography{refs}

\begin{thebibliography}{30}%
\makeatletter
\providecommand \@ifxundefined [1]{%
 \@ifx{#1\undefined}
}%
\providecommand \@ifnum [1]{%
 \ifnum #1\expandafter \@firstoftwo
 \else \expandafter \@secondoftwo
 \fi
}%
\providecommand \@ifx [1]{%
 \ifx #1\expandafter \@firstoftwo
 \else \expandafter \@secondoftwo
 \fi
}%
\providecommand \natexlab [1]{#1}%
\providecommand \enquote  [1]{``#1''}%
\providecommand \bibnamefont  [1]{#1}%
\providecommand \bibfnamefont [1]{#1}%
\providecommand \citenamefont [1]{#1}%
\providecommand \href@noop [0]{\@secondoftwo}%
\providecommand \href [0]{\begingroup \@sanitize@url \@href}%
\providecommand \@href[1]{\@@startlink{#1}\@@href}%
\providecommand \@@href[1]{\endgroup#1\@@endlink}%
\providecommand \@sanitize@url [0]{\catcode `\\12\catcode `\$12\catcode
  `\&12\catcode `\#12\catcode `\^12\catcode `\_12\catcode `\%12\relax}%
\providecommand \@@startlink[1]{}%
\providecommand \@@endlink[0]{}%
\providecommand \url  [0]{\begingroup\@sanitize@url \@url }%
\providecommand \@url [1]{\endgroup\@href {#1}{\urlprefix }}%
\providecommand \urlprefix  [0]{URL }%
\providecommand \Eprint [0]{\href }%
\providecommand \doibase [0]{http://dx.doi.org/}%
\providecommand \selectlanguage [0]{\@gobble}%
\providecommand \bibinfo  [0]{\@secondoftwo}%
\providecommand \bibfield  [0]{\@secondoftwo}%
\providecommand \translation [1]{[#1]}%
\providecommand \BibitemOpen [0]{}%
\providecommand \bibitemStop [0]{}%
\providecommand \bibitemNoStop [0]{.\EOS\space}%
\providecommand \EOS [0]{\spacefactor3000\relax}%
\providecommand \BibitemShut  [1]{\csname bibitem#1\endcsname}%
\let\auto@bib@innerbib\@empty
%</preamble>

\bibitem [{\citenamefont {Lvovsky}\ and\ \citenamefont
  {Raymer}(2009)}]{LvovRaymer09}%
  \BibitemOpen
  \bibfield  {author} {\bibinfo {author} {\bibfnamefont {A.~I.}\ \bibnamefont
  {Lvovsky}}\ and\ \bibinfo {author} {\bibfnamefont {M.}~\bibnamefont
  {Raymer}},\ }\href {\doibase 10.1103/RevModPhys.81.299} {\bibfield  {journal}
  {\bibinfo  {journal} {Reviews of Modern Physics}\ }\textbf {\bibinfo {volume}
  {81}},\ \bibinfo {pages} {299} (\bibinfo {year} {2009})}\BibitemShut
  {NoStop}%

\bibitem [{\citenamefont {Hofheinz}\ \emph {et~al.}(2009)\citenamefont
  {Hofheinz}, \citenamefont {Wang}, \citenamefont {Ansmann}, \citenamefont
  {Bialczak}, \citenamefont {Lucero}, \citenamefont {Neeley}, \citenamefont
  {O'Connell}, \citenamefont {Sank}, \citenamefont {Wenner}, \citenamefont
  {Martinis},\ and\ \citenamefont {Cleland}}]{Hofheinz_09}%
  \BibitemOpen
  \bibfield  {author} {\bibinfo {author} {\bibfnamefont {M.}~\bibnamefont
  {Hofheinz}}, \bibinfo {author} {\bibfnamefont {H.}~\bibnamefont {Wang}},
  \bibinfo {author} {\bibfnamefont {M.}~\bibnamefont {Ansmann}}, \bibinfo
  {author} {\bibfnamefont {R.}~\bibnamefont {Bialczak}}, \bibinfo {author}
  {\bibfnamefont {E.}~\bibnamefont {Lucero}}, \bibinfo {author} {\bibfnamefont
  {M.}~\bibnamefont {Neeley}}, \bibinfo {author} {\bibfnamefont
  {A.}~\bibnamefont {O'Connell}}, \bibinfo {author} {\bibfnamefont
  {D.}~\bibnamefont {Sank}}, \bibinfo {author} {\bibfnamefont {J.}~\bibnamefont
  {Wenner}}, \bibinfo {author} {\bibfnamefont {J.}~\bibnamefont {Martinis}}, \
  and\ \bibinfo {author} {\bibfnamefont {A.}~\bibnamefont {Cleland}},\ }\href
  {\doibase 10.1038/nature08005} {\bibfield  {journal} {\bibinfo  {journal}
  {Nature}\ }\textbf {\bibinfo {volume} {459}},\ \bibinfo {pages} {546}
  (\bibinfo {year} {2009})}\BibitemShut {NoStop}%
\bibitem [{\citenamefont {Leibfried}\ \emph {et~al.}(2005)\citenamefont
  {Leibfried}, \citenamefont {Knill}, \citenamefont {Seidelin}, \citenamefont
  {Britton}, \citenamefont {Blakestad}, \citenamefont {Chiaverini},
  \citenamefont {Hume}, \citenamefont {Itano}, \citenamefont {Jost},
  \citenamefont {Langer}, \citenamefont {Ozeri}, \citenamefont {Reichle},\ and\
  \citenamefont {Wineland}}]{Leibfried_05}%
  \BibitemOpen
  \bibfield  {author} {\bibinfo {author} {\bibfnamefont {D.}~\bibnamefont
  {Leibfried}}, \bibinfo {author} {\bibfnamefont {E.}~\bibnamefont {Knill}},
  \bibinfo {author} {\bibfnamefont {S.}~\bibnamefont {Seidelin}}, \bibinfo
  {author} {\bibfnamefont {J.}~\bibnamefont {Britton}}, \bibinfo {author}
  {\bibfnamefont {R.~B.}\ \bibnamefont {Blakestad}}, \bibinfo {author}
  {\bibfnamefont {J.}~\bibnamefont {Chiaverini}}, \bibinfo {author}
  {\bibfnamefont {D.~B.}\ \bibnamefont {Hume}}, \bibinfo {author}
  {\bibfnamefont {W.~M.}\ \bibnamefont {Itano}}, \bibinfo {author}
  {\bibfnamefont {J.~D.}\ \bibnamefont {Jost}}, \bibinfo {author}
  {\bibfnamefont {C.}~\bibnamefont {Langer}}, \bibinfo {author} {\bibfnamefont
  {R.}~\bibnamefont {Ozeri}}, \bibinfo {author} {\bibfnamefont
  {R.}~\bibnamefont {Reichle}}, \ and\ \bibinfo {author} {\bibfnamefont
  {D.~J.}\ \bibnamefont {Wineland}},\ }\href {\doibase 10.1038/nature04251}
  {\bibfield  {journal} {\bibinfo  {journal} {Nature}\ }\textbf {\bibinfo
  {volume} {438}},\ \bibinfo {pages} {639} (\bibinfo {year}
  {2005})}\BibitemShut {NoStop}%
\bibitem [{\citenamefont {Lvovsky}\ \emph {et~al.}(2009)\citenamefont
  {Lvovsky}, \citenamefont {Tittel},\ and\ \citenamefont
  {Sanders}}]{Lvovsky_09}%
  \BibitemOpen
  \bibfield  {author} {\bibinfo {author} {\bibfnamefont {A.~I.}\ \bibnamefont
  {Lvovsky}}, \bibinfo {author} {\bibfnamefont {W.}~\bibnamefont {Tittel}}, \
  and\ \bibinfo {author} {\bibfnamefont {B.~C.}\ \bibnamefont {Sanders}},\
  }\href@noop {} {\bibfield  {journal} {\bibinfo  {journal} {Nature Photonics}\
  }\textbf {\bibinfo {volume} {3}},\ \bibinfo {pages} {706} (\bibinfo {year}
  {2009})}\BibitemShut {NoStop}%
\bibitem [{\citenamefont {Duan}\ \emph {et~al.}(2001)\citenamefont {Duan},
  \citenamefont {Lukin}, \citenamefont {Cirac},\ and\ \citenamefont
  {Zoller}}]{DLCZ_01}%
  \BibitemOpen
  \bibfield  {author} {\bibinfo {author} {\bibfnamefont {L.~M.}\ \bibnamefont
  {Duan}}, \bibinfo {author} {\bibfnamefont {M.~D.}\ \bibnamefont {Lukin}},
  \bibinfo {author} {\bibfnamefont {J.~I.}\ \bibnamefont {Cirac}}, \ and\
  \bibinfo {author} {\bibfnamefont {P.}~\bibnamefont {Zoller}},\ }\href
  {\doibase 10.1038/35106500} {\bibfield  {journal} {\bibinfo  {journal}
  {Nature}\ }\textbf {\bibinfo {volume} {414}},\ \bibinfo {pages} {413}
  (\bibinfo {year} {2001})}\BibitemShut {NoStop}%
\bibitem [{\citenamefont {You}\ and\ \citenamefont {Chapman}(2000)}]{You_00}%
  \BibitemOpen
  \bibfield  {author} {\bibinfo {author} {\bibfnamefont {L.}~\bibnamefont
  {You}}\ and\ \bibinfo {author} {\bibfnamefont {M.~S.}\ \bibnamefont
  {Chapman}},\ }\href {\doibase 10.1103/PhysRevA.62.052302} {\bibfield
  {journal} {\bibinfo  {journal} {Phys. Rev. A}\ }\textbf {\bibinfo {volume}
  {62}},\ \bibinfo {pages} {052302} (\bibinfo {year} {2000})}\BibitemShut
  {NoStop}%
\bibitem [{\citenamefont {Chiow}\ \emph {et~al.}(2011)\citenamefont {Chiow},
  \citenamefont {Kovachy}, \citenamefont {Chien},\ and\ \citenamefont
  {Kasevich}}]{Kasevich_11}%
  \BibitemOpen
  \bibfield  {author} {\bibinfo {author} {\bibfnamefont {S.-W.}\ \bibnamefont
  {Chiow}}, \bibinfo {author} {\bibfnamefont {T.}~\bibnamefont {Kovachy}},
  \bibinfo {author} {\bibfnamefont {H.-C.}\ \bibnamefont {Chien}}, \ and\
  \bibinfo {author} {\bibfnamefont {M.}~\bibnamefont {Kasevich}},\ }\href
  {\doibase 10.1103/PhysRevLett.107.130403} {\bibfield  {journal} {\bibinfo
  {journal} {Phys. Rev. Lett.}\ }\textbf {\bibinfo {volume} {107}},\ \bibinfo
  {pages} {130403} (\bibinfo {year} {2011})}\BibitemShut {NoStop}%
\bibitem [{\citenamefont {Kitagawa}\ and\ \citenamefont
  {Ueda}(1993)}]{Kitagawa_93}%
  \BibitemOpen
  \bibfield  {author} {\bibinfo {author} {\bibfnamefont {M.}~\bibnamefont
  {Kitagawa}}\ and\ \bibinfo {author} {\bibfnamefont {M.}~\bibnamefont
  {Ueda}},\ }\href {\doibase 10.1103/PhysRevA.47.5138} {\bibfield  {journal}
  {\bibinfo  {journal} {Phys. Rev. A}\ }\textbf {\bibinfo {volume} {47}},\
  \bibinfo {pages} {5138} (\bibinfo {year} {1993})}\BibitemShut {NoStop}%
\bibitem [{\citenamefont {Appel}\ \emph
  {et~al.}(2009{\natexlab{a}})\citenamefont {Appel}, \citenamefont
  {Windpassinger}, \citenamefont {Oblak}, \citenamefont {Hoff}, \citenamefont
  {Kjaergaard},\ and\ \citenamefont {Polzik}}]{Appel_09}%
  \BibitemOpen
  \bibfield  {author} {\bibinfo {author} {\bibfnamefont {J.}~\bibnamefont
  {Appel}}, \bibinfo {author} {\bibfnamefont {P.~J.}\ \bibnamefont
  {Windpassinger}}, \bibinfo {author} {\bibfnamefont {D.}~\bibnamefont
  {Oblak}}, \bibinfo {author} {\bibfnamefont {U.~B.}\ \bibnamefont {Hoff}},
  \bibinfo {author} {\bibfnamefont {N.}~\bibnamefont {Kjaergaard}}, \ and\
  \bibinfo {author} {\bibfnamefont {E.~S.}\ \bibnamefont {Polzik}},\ }\href
  {\doibase 10.1073/pnas.0901550106} {\bibfield  {journal} {\bibinfo  {journal}
  {PNAS}\ }\textbf {\bibinfo {volume} {106}},\ \bibinfo {pages} {10960}
  (\bibinfo {year} {2009}{\natexlab{a}})}\BibitemShut {NoStop}%
\bibitem [{\citenamefont {Leroux}\ \emph {et~al.}(2010)\citenamefont {Leroux},
  \citenamefont {Schleier-Smith},\ and\ \citenamefont
  {Vuleti\ifmmode~\acute{c}\else \'{c}\fi{}}}]{Leroux_10}%
  \BibitemOpen
  \bibfield  {author} {\bibinfo {author} {\bibfnamefont {I.}~\bibnamefont
  {Leroux}}, \bibinfo {author} {\bibfnamefont {M.}~\bibnamefont
  {Schleier-Smith}}, \ and\ \bibinfo {author} {\bibfnamefont {V.}~\bibnamefont
  {Vuleti\ifmmode~\acute{c}\else \'{c}\fi{}}},\ }\href {\doibase
  10.1103/PhysRevLett.104.073602} {\bibfield  {journal} {\bibinfo  {journal}
  {Phys. Rev. Lett.}\ }\textbf {\bibinfo {volume} {104}},\ \bibinfo {pages}
  {073602} (\bibinfo {year} {2010})}\BibitemShut {NoStop}%
\bibitem [{\citenamefont {Chou}\ \emph {et~al.}(2004)\citenamefont {Chou},
  \citenamefont {Polyakov}, \citenamefont {Kuzmich},\ and\ \citenamefont
  {Kimble}}]{Chou_04}%
  \BibitemOpen
  \bibfield  {author} {\bibinfo {author} {\bibfnamefont {C.~W.}\ \bibnamefont
  {Chou}}, \bibinfo {author} {\bibfnamefont {S.~V.}\ \bibnamefont {Polyakov}},
  \bibinfo {author} {\bibfnamefont {A.}~\bibnamefont {Kuzmich}}, \ and\
  \bibinfo {author} {\bibfnamefont {H.~J.}\ \bibnamefont {Kimble}},\ }\href
  {\doibase 10.1103/PhysRevLett.92.213601} {\bibfield  {journal} {\bibinfo
  {journal} {Phys. Rev. Lett.}\ }\textbf {\bibinfo {volume} {92}},\ \bibinfo
  {pages} {213601} (\bibinfo {year} {2004})}\BibitemShut {NoStop}%
\bibitem [{\citenamefont {Thompson}\ \emph {et~al.}(2006)\citenamefont
  {Thompson}, \citenamefont {Simon}, \citenamefont {Loh},\ and\ \citenamefont
  {Vuleti\`{c}}}]{Thompson_06}%
  \BibitemOpen
  \bibfield  {author} {\bibinfo {author} {\bibfnamefont {J.}~\bibnamefont
  {Thompson}}, \bibinfo {author} {\bibfnamefont {J.}~\bibnamefont {Simon}},
  \bibinfo {author} {\bibfnamefont {H.}~\bibnamefont {Loh}}, \ and\ \bibinfo
  {author} {\bibfnamefont {V.}~\bibnamefont {Vuleti\`{c}}},\ }\href {\doibase
  10.1126/science.1127676} {\bibfield  {journal} {\bibinfo  {journal}
  {Science}\ }\textbf {\bibinfo {volume} {313}},\ \bibinfo {pages} {74}
  (\bibinfo {year} {2006})},\ \Eprint
  {http://arxiv.org/abs/http://www.sciencemag.org/content/313/5783/74.full.pdf}
  {http://www.sciencemag.org/content/313/5783/74.full.pdf} \BibitemShut
  {NoStop}%
\bibitem [{\citenamefont {Bimbard}\ \emph {et~al.}(2010)\citenamefont
  {Bimbard}, \citenamefont {Jain}, \citenamefont {MacRae},\ and\ \citenamefont
  {Lvovsky}}]{Bimbard_10}%
  \BibitemOpen
  \bibfield  {author} {\bibinfo {author} {\bibfnamefont {E.}~\bibnamefont
  {Bimbard}}, \bibinfo {author} {\bibfnamefont {N.}~\bibnamefont {Jain}},
  \bibinfo {author} {\bibfnamefont {A.}~\bibnamefont {MacRae}}, \ and\ \bibinfo
  {author} {\bibfnamefont {A.~I.}\ \bibnamefont {Lvovsky}},\ }\href {\doibase
  10.1038/NPHOTON.2010.6} {\bibfield  {journal} {\bibinfo  {journal} {Nature
  Photonics}\ }\textbf {\bibinfo {volume} {4}},\ \bibinfo {pages} {243}
  (\bibinfo {year} {2010})}\BibitemShut {NoStop}%
\bibitem [{\citenamefont {Fleischhauer}\ and\ \citenamefont
  {Lukin}(2000)}]{DSP_00}%
  \BibitemOpen
  \bibfield  {author} {\bibinfo {author} {\bibfnamefont {M.}~\bibnamefont
  {Fleischhauer}}\ and\ \bibinfo {author} {\bibfnamefont {M.~D.}\ \bibnamefont
  {Lukin}},\ }\href {\doibase 10.1103/PhysRevLett.84.5094} {\bibfield
  {journal} {\bibinfo  {journal} {Phys. Rev. Lett.}\ }\textbf {\bibinfo
  {volume} {84}},\ \bibinfo {pages} {5094} (\bibinfo {year}
  {2000})}\BibitemShut {NoStop}%
\bibitem [{\citenamefont {Fernholz}\ \emph {et~al.}(2008)\citenamefont
  {Fernholz}, \citenamefont {Krauter}, \citenamefont {Jensen}, \citenamefont
  {Sherson}, \citenamefont {S\o{}rensen},\ and\ \citenamefont
  {Polzik}}]{Fernholz_08}%
  \BibitemOpen
  \bibfield  {author} {\bibinfo {author} {\bibfnamefont {T.}~\bibnamefont
  {Fernholz}}, \bibinfo {author} {\bibfnamefont {H.}~\bibnamefont {Krauter}},
  \bibinfo {author} {\bibfnamefont {K.}~\bibnamefont {Jensen}}, \bibinfo
  {author} {\bibfnamefont {J.~F.}\ \bibnamefont {Sherson}}, \bibinfo {author}
  {\bibfnamefont {A.~S.}\ \bibnamefont {S\o{}rensen}}, \ and\ \bibinfo {author}
  {\bibfnamefont {E.~S.}\ \bibnamefont {Polzik}},\ }\href {\doibase
  10.1103/PhysRevLett.101.073601} {\bibfield  {journal} {\bibinfo  {journal}
  {Phys. Rev. Lett.}\ }\textbf {\bibinfo {volume} {101}},\ \bibinfo {pages}
  {073601} (\bibinfo {year} {2008})}\BibitemShut {NoStop}%
\bibitem [{\citenamefont {Neergaard-Nielsen}\ \emph {et~al.}(2007)\citenamefont
  {Neergaard-Nielsen}, \citenamefont {Nielsen}, \citenamefont {Takahashi},
  \citenamefont {Vistnes},\ and\ \citenamefont {Polzik}}]{Polzik_07}%
  \BibitemOpen
  \bibfield  {author} {\bibinfo {author} {\bibfnamefont {J.~S.}\ \bibnamefont
  {Neergaard-Nielsen}}, \bibinfo {author} {\bibfnamefont {B.~M.}\ \bibnamefont
  {Nielsen}}, \bibinfo {author} {\bibfnamefont {H.}~\bibnamefont {Takahashi}},
  \bibinfo {author} {\bibfnamefont {A.~I.}\ \bibnamefont {Vistnes}}, \ and\
  \bibinfo {author} {\bibfnamefont {E.~S.}\ \bibnamefont {Polzik}},\ }\href
  {http://www.ncbi.nlm.nih.gov/pubmed/19547121} {\bibfield  {journal} {\bibinfo
   {journal} {Optics Express}\ }\textbf {\bibinfo {volume} {15}},\ \bibinfo
  {pages} {7940} (\bibinfo {year} {2007})}\BibitemShut {NoStop}%
\bibitem [{\citenamefont {Halder}\ \emph {et~al.}(2008)\citenamefont {Halder},
  \citenamefont {Beveratos}, \citenamefont {Thew}, \citenamefont {Jorel},
  \citenamefont {Zbinden},\ and\ \citenamefont {Gisin}}]{Halder_08}%
  \BibitemOpen
  \bibfield  {author} {\bibinfo {author} {\bibfnamefont {M.}~\bibnamefont
  {Halder}}, \bibinfo {author} {\bibfnamefont {A.}~\bibnamefont {Beveratos}},
  \bibinfo {author} {\bibfnamefont {R.}~\bibnamefont {Thew}}, \bibinfo {author}
  {\bibfnamefont {C.}~\bibnamefont {Jorel}}, \bibinfo {author} {\bibfnamefont
  {H.}~\bibnamefont {Zbinden}}, \ and\ \bibinfo {author} {\bibfnamefont
  {N.}~\bibnamefont {Gisin}},\ }\href
  {http://stacks.iop.org/1367-2630/10/i=2/a=023027} {\bibfield  {journal}
  {\bibinfo  {journal} {New Journal of Physics}\ }\textbf {\bibinfo {volume}
  {10}},\ \bibinfo {pages} {023027} (\bibinfo {year} {2008})}\BibitemShut
  {NoStop}%
\bibitem [{\citenamefont {Specht}\ \emph {et~al.}(2009)\citenamefont {Specht},
  \citenamefont {Bochmann}, \citenamefont {MŸcke}, \citenamefont {Weber},
  \citenamefont {Figueroa}, \citenamefont {Moehring},\ and\ \citenamefont
  {Rempe}}]{Specht_09}%
  \BibitemOpen
  \bibfield  {author} {\bibinfo {author} {\bibfnamefont {H.~P.}\ \bibnamefont
  {Specht}}, \bibinfo {author} {\bibfnamefont {J.}~\bibnamefont {Bochmann}},
  \bibinfo {author} {\bibfnamefont {M.}~\bibnamefont {MŸcke}}, \bibinfo
  {author} {\bibfnamefont {B.}~\bibnamefont {Weber}}, \bibinfo {author}
  {\bibfnamefont {E.}~\bibnamefont {Figueroa}}, \bibinfo {author}
  {\bibfnamefont {D.~L.}\ \bibnamefont {Moehring}}, \ and\ \bibinfo {author}
  {\bibfnamefont {G.}~\bibnamefont {Rempe}},\ }\href {\doibase
  doi:10.1038/nphoton.2009.115} {\bibfield  {journal} {\bibinfo  {journal}
  {Nature Photonics}\ }\textbf {\bibinfo {volume} {3}},\ \bibinfo {pages} {469
  } (\bibinfo {year} {2009})}\BibitemShut {NoStop}%
\bibitem [{\citenamefont {Hostein}\ \emph {et~al.}(2009)\citenamefont
  {Hostein}, \citenamefont {Braive}, \citenamefont {Larque}, \citenamefont
  {Lee}, \citenamefont {Talneau}, \citenamefont {Gratiet}, \citenamefont
  {Robert-Philip}, \citenamefont {Sagnes},\ and\ \citenamefont
  {Beveratos}}]{Hostein_09}%
  \BibitemOpen
  \bibfield  {author} {\bibinfo {author} {\bibfnamefont {R.}~\bibnamefont
  {Hostein}}, \bibinfo {author} {\bibfnamefont {R.}~\bibnamefont {Braive}},
  \bibinfo {author} {\bibfnamefont {M.}~\bibnamefont {Larque}}, \bibinfo
  {author} {\bibfnamefont {K.-H.}\ \bibnamefont {Lee}}, \bibinfo {author}
  {\bibfnamefont {A.}~\bibnamefont {Talneau}}, \bibinfo {author} {\bibfnamefont
  {L.~L.}\ \bibnamefont {Gratiet}}, \bibinfo {author} {\bibfnamefont
  {I.}~\bibnamefont {Robert-Philip}}, \bibinfo {author} {\bibfnamefont
  {I.}~\bibnamefont {Sagnes}}, \ and\ \bibinfo {author} {\bibfnamefont
  {A.}~\bibnamefont {Beveratos}},\ }\href {\doibase 10.1063/1.3104855}
  {\bibfield  {journal} {\bibinfo  {journal} {Applied Physics Letters}\
  }\textbf {\bibinfo {volume} {94}},\ \bibinfo {eid} {123101} (\bibinfo {year}
  {2009})}\BibitemShut {NoStop}%
\bibitem [{\citenamefont {Batalov}\ \emph {et~al.}(2008)\citenamefont
  {Batalov}, \citenamefont {Zierl}, \citenamefont {Gaebel}, \citenamefont
  {Neumann}, \citenamefont {Chan}, \citenamefont {Balasubramanian},
  \citenamefont {Hemmer}, \citenamefont {Jelezko},\ and\ \citenamefont
  {Wrachtrup}}]{Batalov_08}%
  \BibitemOpen
  \bibfield  {author} {\bibinfo {author} {\bibfnamefont {A.}~\bibnamefont
  {Batalov}}, \bibinfo {author} {\bibfnamefont {C.}~\bibnamefont {Zierl}},
  \bibinfo {author} {\bibfnamefont {T.}~\bibnamefont {Gaebel}}, \bibinfo
  {author} {\bibfnamefont {P.}~\bibnamefont {Neumann}}, \bibinfo {author}
  {\bibfnamefont {I.-Y.}\ \bibnamefont {Chan}}, \bibinfo {author}
  {\bibfnamefont {G.}~\bibnamefont {Balasubramanian}}, \bibinfo {author}
  {\bibfnamefont {P.~R.}\ \bibnamefont {Hemmer}}, \bibinfo {author}
  {\bibfnamefont {F.}~\bibnamefont {Jelezko}}, \ and\ \bibinfo {author}
  {\bibfnamefont {J.}~\bibnamefont {Wrachtrup}},\ }\href {\doibase
  10.1103/PhysRevLett.100.077401} {\bibfield  {journal} {\bibinfo  {journal}
  {Phys. Rev. Lett.}\ }\textbf {\bibinfo {volume} {100}},\ \bibinfo {pages}
  {077401} (\bibinfo {year} {2008})}\BibitemShut {NoStop}%
\bibitem [{\citenamefont {McCormick}\ \emph {et~al.}(2007)\citenamefont
  {McCormick}, \citenamefont {Boyer}, \citenamefont {Arimondo},\ and\
  \citenamefont {Lett}}]{McCormick_07}%
  \BibitemOpen
  \bibfield  {author} {\bibinfo {author} {\bibfnamefont {C.~F.}\ \bibnamefont
  {McCormick}}, \bibinfo {author} {\bibfnamefont {V.}~\bibnamefont {Boyer}},
  \bibinfo {author} {\bibfnamefont {E.}~\bibnamefont {Arimondo}}, \ and\
  \bibinfo {author} {\bibfnamefont {P.~D.}\ \bibnamefont {Lett}},\ }\href
  {http://www.ncbi.nlm.nih.gov/pubmed/17186056} {\bibfield  {journal} {\bibinfo
   {journal} {Optics Letters}\ }\textbf {\bibinfo {volume} {32}},\ \bibinfo
  {pages} {178} (\bibinfo {year} {2007})}\BibitemShut {NoStop}%
\bibitem [{\citenamefont {Boyer}\ \emph {et~al.}(2008)\citenamefont {Boyer},
  \citenamefont {Marino}, \citenamefont {Pooser},\ and\ \citenamefont
  {Lett}}]{Boyer_08}%
  \BibitemOpen
  \bibfield  {author} {\bibinfo {author} {\bibfnamefont {V.}~\bibnamefont
  {Boyer}}, \bibinfo {author} {\bibfnamefont {A.}~\bibnamefont {Marino}},
  \bibinfo {author} {\bibfnamefont {R.}~\bibnamefont {Pooser}}, \ and\ \bibinfo
  {author} {\bibfnamefont {P.}~\bibnamefont {Lett}},\ }\href {\doibase
  10.1126/science.1158275} {\bibfield  {journal} {\bibinfo  {journal}
  {Science}\ }\textbf {\bibinfo {volume} {321}},\ \bibinfo {pages} {544}
  (\bibinfo {year} {2008})}\BibitemShut {NoStop}%
\bibitem [{\citenamefont {Palittapongarnpim}\ \emph {et~al.}(2012)\citenamefont
  {Palittapongarnpim}, \citenamefont {MacRae},\ and\ \citenamefont
  {Lvovsky}}]{Cavity_Paper}%
  \BibitemOpen
  \bibfield  {author} {\bibinfo {author} {\bibfnamefont {P.}~\bibnamefont
  {Palittapongarnpim}}, \bibinfo {author} {\bibfnamefont {A.}~\bibnamefont
  {MacRae}}, \ and\ \bibinfo {author} {\bibfnamefont {A.~I.}\ \bibnamefont
  {Lvovsky}},\ }\href@noop {} {\bibfield  {journal} {\bibinfo  {journal} {ArXiv
  [physics.ins-det]}\ }\textbf {\bibinfo {volume} {1203.4843v1}} (\bibinfo
  {year} {2012})}\BibitemShut {NoStop}%
\bibitem [{\citenamefont {Boyd}(2008)}]{Boyd}%
  \BibitemOpen
  \bibfield  {author} {\bibinfo {author} {\bibfnamefont {R.}~\bibnamefont
  {Boyd}},\ }in\ \href@noop {} {\emph {\bibinfo {booktitle} {Nonlinear
  Optics}}}\ (\bibinfo  {publisher} {Academic Press},\ \bibinfo {year} {2008})\
  pp.\ \bibinfo {pages} {473--508}\BibitemShut {NoStop}%
\bibitem [{\citenamefont {Kumar}\ \emph {et~al.}(2011)\citenamefont {Kumar},
  \citenamefont {Barrios}, \citenamefont {MacRae}, \citenamefont {Cairns},
  \citenamefont {Huntington},\ and\ \citenamefont {Lvovsky}}]{Kumar_11}%
  \BibitemOpen
  \bibfield  {author} {\bibinfo {author} {\bibfnamefont {R.}~\bibnamefont
  {Kumar}}, \bibinfo {author} {\bibfnamefont {E.}~\bibnamefont {Barrios}},
  \bibinfo {author} {\bibfnamefont {A.}~\bibnamefont {MacRae}}, \bibinfo
  {author} {\bibfnamefont {E.}~\bibnamefont {Cairns}}, \bibinfo {author}
  {\bibfnamefont {E.}~\bibnamefont {Huntington}}, \ and\ \bibinfo {author}
  {\bibfnamefont {A.~I.}\ \bibnamefont {Lvovsky}},\ }\href@noop {} {\bibfield
  {journal} {\bibinfo  {journal} {ArXiv [physics.ins-det]}\ }\textbf {\bibinfo
  {volume} {1111.4012v1}} (\bibinfo {year} {2011})}\BibitemShut {NoStop}%
\bibitem [{\citenamefont {Appel}\ \emph
  {et~al.}(2009{\natexlab{b}})\citenamefont {Appel}, \citenamefont {MacRae},\
  and\ \citenamefont {Lvovsky}}]{PLL_09}%
  \BibitemOpen
  \bibfield  {author} {\bibinfo {author} {\bibfnamefont {J.}~\bibnamefont
  {Appel}}, \bibinfo {author} {\bibfnamefont {A.}~\bibnamefont {MacRae}}, \
  and\ \bibinfo {author} {\bibfnamefont {A.~I.}\ \bibnamefont {Lvovsky}},\
  }\href {\doibase 10.1088/0957-0233/20/5/055302} {\bibfield  {journal}
  {\bibinfo  {journal} {Measurement Science and Technology}\ }\textbf {\bibinfo
  {volume} {20}},\ \bibinfo {pages} {055302} (\bibinfo {year}
  {2009}{\natexlab{b}})}\BibitemShut {NoStop}%
\bibitem [{\citenamefont {{T. Aichele}}\ \emph {et~al.}(2002)\citenamefont {{T.
  Aichele}}, \citenamefont {{A.~I.~Lvovsky}},\ and\ \citenamefont
  {{S.~Schiller}}}]{Aichele_02}%
  \BibitemOpen
  \bibfield  {author} {\bibinfo {author} {\bibnamefont {{T. Aichele}}},
  \bibinfo {author} {\bibnamefont {{A.~I.~Lvovsky}}}, \ and\ \bibinfo {author}
  {\bibnamefont {{S.~Schiller}}},\ }\href {\doibase 10.1140/epjd/e20020028}
  {\bibfield  {journal} {\bibinfo  {journal} {Eur. Phys. J. D}\ }\textbf
  {\bibinfo {volume} {18}},\ \bibinfo {pages} {237} (\bibinfo {year}
  {2002})}\BibitemShut {NoStop}%
\bibitem [{\citenamefont {Lvovsky}(2004)}]{MaxLik_04}%
  \BibitemOpen
  \bibfield  {author} {\bibinfo {author} {\bibfnamefont {A.~I.}\ \bibnamefont
  {Lvovsky}},\ }\href {http://stacks.iop.org/1464-4266/6/i=6/a=014} {\bibfield
  {journal} {\bibinfo  {journal} {Journal of Optics B: Quantum and
  Semiclassical Optics}\ }\textbf {\bibinfo {volume} {6}},\ \bibinfo {pages}
  {S556} (\bibinfo {year} {2004})}\BibitemShut {NoStop}%
\bibitem [{\citenamefont {\ifmmode \check{R}\else
  \v{R}\fi{}eh\'a\ifmmode~\check{c}\else \v{c}\fi{}ek}\ \emph
  {et~al.}(2007)\citenamefont {\ifmmode \check{R}\else
  \v{R}\fi{}eh\'a\ifmmode~\check{c}\else \v{c}\fi{}ek}, \citenamefont {Hradil},
  \citenamefont {Knill},\ and\ \citenamefont {Lvovsky}}]{MaxLik_07}%
  \BibitemOpen
  \bibfield  {author} {\bibinfo {author} {\bibfnamefont {J.}~\bibnamefont
  {\ifmmode \check{R}\else \v{R}\fi{}eh\'a\ifmmode~\check{c}\else
  \v{c}\fi{}ek}}, \bibinfo {author} {\bibfnamefont {Z.}~\bibnamefont {Hradil}},
  \bibinfo {author} {\bibfnamefont {E.}~\bibnamefont {Knill}}, \ and\ \bibinfo
  {author} {\bibfnamefont {A.~I.}\ \bibnamefont {Lvovsky}},\ }\href {\doibase
  10.1103/PhysRevA.75.042108} {\bibfield  {journal} {\bibinfo  {journal} {Phys.
  Rev. A}\ }\textbf {\bibinfo {volume} {75}},\ \bibinfo {pages} {042108}
  (\bibinfo {year} {2007})}\BibitemShut {NoStop}%
\bibitem [{\citenamefont {Lvovsky}\ \emph {et~al.}(2001)\citenamefont
  {Lvovsky}, \citenamefont {Hansen}, \citenamefont {Aichele}, \citenamefont
  {Benson}, \citenamefont {Mlynek},\ and\ \citenamefont
  {Schiller}}]{Lvovsky_01}%
  \BibitemOpen
  \bibfield  {author} {\bibinfo {author} {\bibfnamefont {A.~I.}\ \bibnamefont
  {Lvovsky}}, \bibinfo {author} {\bibfnamefont {H.}~\bibnamefont {Hansen}},
  \bibinfo {author} {\bibfnamefont {T.}~\bibnamefont {Aichele}}, \bibinfo
  {author} {\bibfnamefont {O.}~\bibnamefont {Benson}}, \bibinfo {author}
  {\bibfnamefont {J.}~\bibnamefont {Mlynek}}, \ and\ \bibinfo {author}
  {\bibfnamefont {S.}~\bibnamefont {Schiller}},\ }\href {\doibase
  10.1103/PhysRevLett.87.050402} {\bibfield  {journal} {\bibinfo  {journal}
  {Phys. Rev. Lett.}\ }\textbf {\bibinfo {volume} {87}},\ \bibinfo {pages}
  {050402} (\bibinfo {year} {2001})}\BibitemShut {NoStop}%
\end{thebibliography}


%

\end{document}